\documentclass[bibyear,letter]{aa}
\usepackage{graphicx}
\usepackage[varg]{txfonts}
\usepackage{placeins}
\usepackage{wrapfig}
\usepackage{color}

\begin{document}

\title{
  Contemporaneous 
  high-angular-resolution imaging of the AGB star W Hya in
  vibrationally excited \mbox{H$_2$O}\ lines and visible polarized light 
  with ALMA and VLT/SPHERE-ZIMPOL
}

\author{K.~Ohnaka\inst{1} 
\and
K.~T.~Wong\inst{2}
\and
G.~Weigelt\inst{3} 
\and
K.-H.~Hofmann\inst{3} 
}

\offprints{K.~Ohnaka}

\institute{
  Instituto de Astrof\'isica, Departamento de F\'isica y Astronom\'ia,
  Facultad de Ciencias Exactas, 
  Universidad Andr\'es Bello,
Fern\'andez Concha 700, Las Condes, Santiago, Chile\\
\email{k1.ohnaka@gmail.com}
\and
Department of Physics and Astronomy, Uppsala University,
Box 516, 751 20 Uppsala, Sweden
\and
Max-Planck-Institut f\"{u}r Radioastronomie, 
Auf dem H\"{u}gel 69, 53121 Bonn, Germany
}

\date{Received / Accepted }

\abstract
{}
{
We present contemporaneous high-angular-resolution millimeter imaging 
and visible polarimetric imaging of the nearby asymptotic giant branch (AGB) star
W~Hya to better understand the dynamics and dust formation within a few 
stellar radii. 
}
{
  The star 
  \mbox{W~Hya}\ was observed in two vibrationally excited \mbox{H$_2$O}\ lines
  at 268 and 251~GHz with Atacama Large Millimeter/submillimeter Array 
  (ALMA) at a spatial resolution of 16$\times$20~mas and
  at 748 and 820~nm at a resolution of 26$\times$27~mas with
  the Very Large Telescope (VLT)/Spectro-Polarimetric High-contrast Exoplanet
  REsearch (SPHERE)-Zurich Imaging Polarimeter (ZIMPOL). 
}
{
ALMA's high spatial resolution allowed us to image 
strong emission of the vibrationally excited
\mbox{H$_2$O}\ line at 268~GHz ($\varv_2$ = 2, $J_{K_a,K_c}$ = $6_{5,2}$--$7_{4,3}$)
over the stellar surface instead of absorption against the continuum, which is 
expected for thermal excitation.
Strong, spotty emission was also detected along and just outside the
stellar disk limb at an angular distance of $\sim$40~mas ($\sim$1.9~\mbox{$R_{\star}$}),
extending to $\sim$60~mas ($\sim$2.9~\mbox{$R_{\star}$}). 
Another \mbox{H$_2$O}\ line ($\varv_2$ = 2, $J_{K_a,K_c}$ = $9_{2,8}$--$8_{3,5}$) at
251~GHz with a similar upper-level energy was tentatively identified, which 
shows absorption over the stellar surface. 
This suggests that the emission over the surface seen in the 268~GHz \mbox{H$_2$O}\
line is suprathermal or even maser emission. 
The estimated gas temperature and \mbox{H$_2$O}\ density are consistent with
the radiatively pumped masers.
The 268~GHz \mbox{H$_2$O}\ line reveals global infall at up to $\sim$15~\mbox{km s$^{-1}$}\ within
2--3~\mbox{$R_{\star}$}, but outflows at up to $\sim$8~\mbox{km s$^{-1}$}\
are also present. 
The polarized intensity maps obtained in the visible reveal clumpy dust 
clouds forming within $\sim$40~mas ($\sim$1.9~\mbox{$R_{\star}$}) with a particularly
prominent cloud in the SW quadrant and a weaker cloud in the
east. 
The 268~GHz \mbox{H$_2$O}\ emission overlaps very well with the visible polarized
intensity maps, which suggests  that both
the nonthermal and likely maser \mbox{H$_2$O}\ emission and the dust originate from dense, cool pockets in the inhomogeneous atmosphere
within $\sim$2--3~\mbox{$R_{\star}$}. 
}
{}

\keywords{
infrared: stars --
techniques: interferometric -- 
stars: imaging -- 
stars: mass-loss -- 
stars: AGB and post-AGB --
(Stars:) circumstellar matter
}   

\titlerunning{Contemporaneous high angular resolution millimeter and
  visible imaging of the AGB star W~Hya}
\authorrunning{Ohnaka et al.}
\maketitle

\section{Introduction}
\label{sect_intro}

Mass loss on the asymptotic giant branch (AGB) plays an important role
not only in stellar evolution but also in the chemical evolution of galaxies.
The initial--final mass relation of low- and intermediate-mass stars
shows that stars can lose up to $\sim$80\% of their initial mass before
they evolve to white dwarfs (Cummings et al. \cite{cummings18}). 
Despite such importance, the mass-loss mechanism in AGB stars is not yet
fully understood. 
According to the pulsation-enhanced dust-driven outflow scenario, 
large-amplitude stellar 
pulsation levitates the material, which leads to density enhancement at
the cool, upper atmosphere, where dust can form, and the radiation pressure
on the dust grains can drive the mass loss 
(H\"ofner \& Olofsson \cite{hoefner18}).
The recent 3D models of AGB stars show that
convection can also levitate the material, which leads to 
clumpy stellar winds (Freytag \& H\"ofner \cite{freytag23}).

To clarify the long-standing problem of the AGB mass loss, 
it is indispensable to probe the region within $\sim$10~\mbox{$R_{\star}$}, 
where dust forms and the wind accelerates. 
The oxygen-rich AGB star W~Hya has been studied with various observational
techniques from the visible to the infrared to the radio 
(e.g., Zhao-Geisler et al. \cite{zhao-geisler11}; 
Khouri et al. \cite{khouri15} and references therein)
because of its proximity 
($98^{+30}_{-18}$~pc, Vlemmings et al. \cite{vlemmings03}).
The light curve shows clear periodicity (Woodruff et al. \cite{woodruff08})
with a period of 389 days (Uttenthaler et al. \cite{uttenthaler11}). 
Woodruff et al. (\cite{woodruff09}) show that 
the minimum near-infrared uniform-disk (UD) diameter of W Hya at phase 0.58
(near minimum light) is $\sim$36 mas at 1.2--1.3~\mbox{$\mu$m}.
This is considered
to correspond to the star's angular diameter as seen in the deepest photospheric
layers. 
We converted the UD diameter to the fully darkened disk diameter
by multiplying it by the factor 1.15 derived by Zhao-Geisler et al.
(\cite{zhao-geisler11}). 
This resulted in a stellar angular diameter of 41.4~mas 
(i.e., 20.7~mas as the star's angular radius \mbox{$R_{\star}$}). 

Taking advantage of visible polarimetric imaging with the
Spectro-Polarimetric High-contrast Exoplanet
REsearch (SPHERE)-Zurich Imaging Polarimeter (ZIMPOL) instrument
(Beuzit et al. \cite{beuzit08}; Schmid et al. \cite{schmid18}) at the 
Very Large Telescope (VLT), 
Ohnaka et al. (\cite{ohnaka16}) spatially resolved the clumpy dust clouds
forming around W~Hya at an angular distance of 
$\sim$50~mas ($\sim$2.4~\mbox{$R_{\star}$}), 
which is similar to the angular distance derived by
Norris et al. (\cite{norris12}) based on polarimetric aperture-masking
experiments.
The modeling of Norris et al. (\cite{norris12}) and
Ohnaka et al. (\cite{ohnaka16}) 
suggests a predominance of large (0.3--0.5~\mbox{$\mu$m}), transparent grains 
of \mbox{Al$_2$O$_3$}, \mbox{Mg$_2$SiO$_4$}, or \mbox{MgSiO$_3$}\ in the clumpy clouds, 
lending support to the scenario of scattering-driven mass loss
(H\"ofner \cite{hoefner08}). 
Furthermore, the second-epoch
ZIMPOL 
observations by Ohnaka et al. (\cite{ohnaka17}) revealed the morphological
change in the dust clouds as well as a change in the grain size from 
0.4--0.5~\mbox{$\mu$m}\ to $\sim$0.1~\mbox{$\mu$m}. 
Similar clumpy dust cloud formation has been imaged in other nearby AGB
stars, R~Dor, IK~Tau, and $o$~Cet (Khouri et al. \cite{khouri16},
\cite{khouri18}; Adam \& Ohnaka \cite{adam19}). 
Using the Atacama Large Millimeter/submillimeter Array (ALMA), 
Takigawa et al. (\cite{takigawa17}) imaged \mbox{W~Hya}\ in 
the gas-phase AlO line at 344~GHz at an epoch 
between two ZIMPOL observations of Ohnaka et al.
(\cite{ohnaka16}, \cite{ohnaka17}). Takigawa et al. (\cite{takigawa17}) found that 
the distribution of AlO is in 
excellent agreement with the clumpy dust clouds seen in the
polarized intensity maps. In this Letter, we present contemporaneous 
high-angular-resolution imaging of \mbox{W~Hya}\ in two vibrationally excited \mbox{H$_2$O}\
lines with ALMA and visible polarimetric imaging with VLT/SPHERE-ZIMPOL.

\begin{figure*}
\begin{center}
\resizebox{\hsize}{!}{\rotatebox{0}{\includegraphics{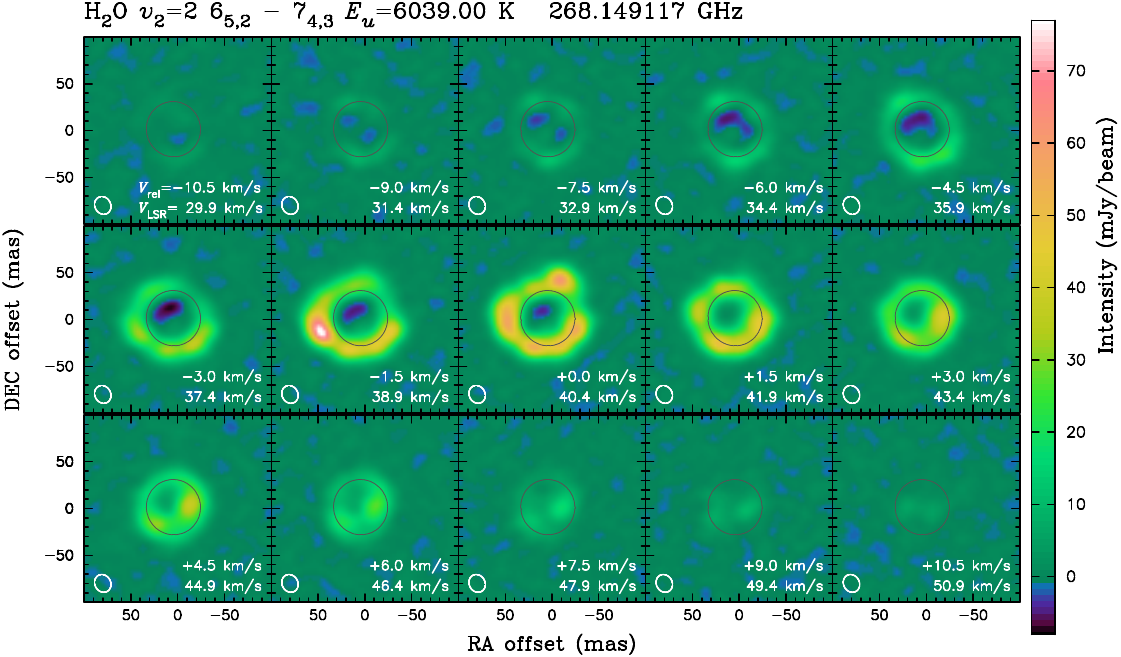}}}
\end{center}
\caption{
  Continuum-subtracted channel maps of \mbox{W~Hya}\ obtained in the vibrationally
  excited \mbox{H$_2$O}\ maser line 
  ($\varv_2$ = 2, $J_{K_a,K_c}$ = $6_{5,2}$--$7_{4,3}$, $E_u$ = 6039.0~K)
  at 268.149117~GHz.
  The gray circles represent the ellipse fitted to the millimeter continuum
  image.
  In the lower right corner of each panel, 
  the LSR velocity and the relative
  velocity ($\mbox{$V_{\rm rel}$}\ = \mbox{$V_{\rm LSR}$} - \mbox{$V_{\rm sys}$}$, $\mbox{$V_{\rm sys}$} = 40.4$~\mbox{km s$^{-1}$}) are shown.
  The restoring beam size 
  is shown in the lower left corner of each panel.
  North is up, and east to the left. 
}
\label{channelmap_h2o_268}
\end{figure*}

\begin{figure*}
\begin{center}
\resizebox{\hsize}{!}{\rotatebox{0}{\includegraphics{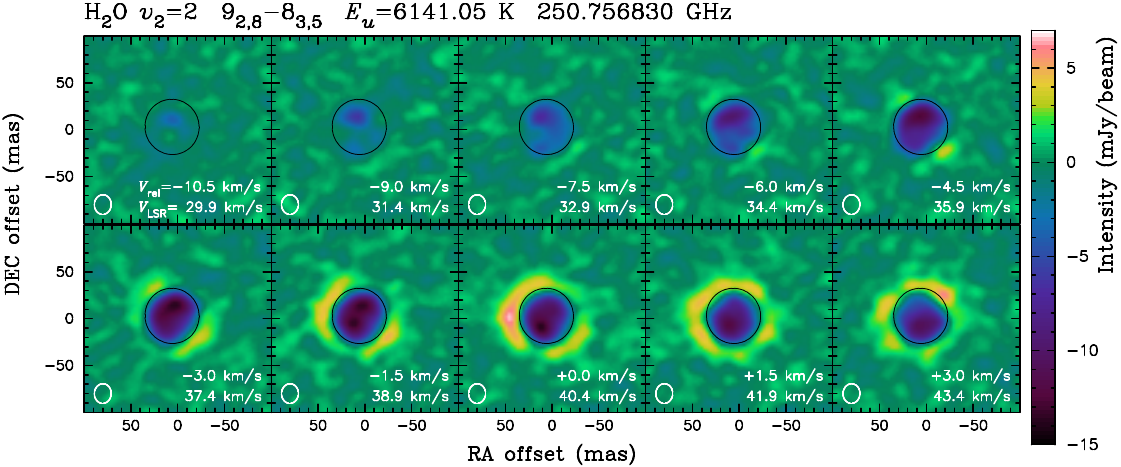}}}
\end{center}
\caption{
  Continuum-subtracted channel maps of \mbox{W~Hya}\ obtained for the vibrationally
  excited \mbox{H$_2$O}\ line 
  ($\varv_2$ = 2, $J_{K_a,K_c}$ = $9_{2,8}$--$8_{3,5}$, $E_u = 6141.05$~K) at 
  250.756830~GHz, shown in the same manner as
  Fig.~\ref{channelmap_h2o_268}.
  The channel maps at the velocities more redshifted than 
  \mbox{$V_{\rm rel}$}\ = 3.0~\mbox{km s$^{-1}$}\ are not shown, because the signals are affected by 
  the adjacent \mbox{Si$^{17}$O}\ line ($\varv=0$, $J$ = 6 -- 5) at 250.744695~GHz.
}
\label{channelmap_h2o_250}
\end{figure*}

\section{Observations and data reduction}
\label{sect_obs}

We observed \mbox{W~Hya}\ with ALMA on 2019 June 8 (UTC) using the C43-10 
configuration (Program ID: 2018.1.01239.S, P.I.: K.~Ohnaka), 
when the variability phase of \mbox{W~Hya}\ was 0.53 at minimum light. 
The shortest and longest baselines were 83.1~m and 16.2~km, respectively. 
The largest recoverable scale is 190~mas.
We observed in nine spectral windows between 250.7 and 268.2~GHz. 
Two vibrationally excited \mbox{H$_2$O}\ lines reported in this {Letter} are
located in the spectral windows centered at 250.727199 and 268.168335~GHz
with bandwidths of 468.75 and 937.50~MHz observed at velocity resolutions of
1.17 and 1.09~\mbox{km s$^{-1}$}, respectively. 

The ALMA data were reduced with the Common Astronomy Software Applications
(CASA version 5.6.1-8; The CASA Team \cite{casa22}), following standard
calibration procedures with the ALMA pipeline. We also self-calibrated
the data with the continuum of W~Hya. The image reconstruction was carried out
with the task tclean with a robust parameter of 0.5.
In total, we identified nearly 60 molecular lines, 
which include \mbox{H$_2$O}, \mbox{$^{29}$SiO}, \mbox{$^{30}$SiO}, \mbox{Si$^{17}$O}, OH, \mbox{SO$_{2}$}, \mbox{$^{34}$SO$_{2}$}, SO$^{18}$O,
AlO, AlOH, TiO, $^{49}$TiO, $^{50}$TiO, \mbox{TiO$_{2}$}, SO, and HCN. 
The complete results will be presented in a forthcoming paper
(Ohnaka et al., in prep). 
The images of the \mbox{H$_2$O}\ lines at 251 and 268~GHz were obtained 
with a restoring beam of 21$\times$18~mas and 20$\times$17~mas, respectively.
The root mean square (RMS) noise is estimated to be 0.56~mJy/beam in both
spectral windows.

We also used 
archival polarimetric imaging data of \mbox{W~Hya}\ taken at 748 and 820~nm 
on 2019 May 30 (UTC), 
just nine days before our ALMA observations, using 
VLT/SPHERE-ZIMPOL as part of the program 0103.D-0168 (P.I.:
T.~Khouri). 
The spatial resolution
estimated from the 2D Gaussian fit to the point spread function (PSF)
reference star is 26$\times$27~mas. The observations and data reduction
are described in Appendix~\ref{appendix_zimpol}.

\section{Vibrationally excited \mbox{H$_2$O}\ line at 268~GHz ($\varv_2 = 2$,
  $6_{5,2}$ -- $7_{4,3}$) }
\label{sect_res_h2o_268}

Figure~\ref{cont_image} shows the continuum image 
obtained in the 268.2~GHz spectral window. 
Fitting the continuum image with a uniform elliptical disk results
in a major and minor axis of $59.1 \pm 0.3$~mas and $57.7 \pm 0.2$~mas
($\sim$1.4 times larger than our adopted stellar
angular diameter, which was also obtained near minimum light), 
with the position angle of the major axis being 16 $\pm$ 23\degr. 
  The uniform elliptical disk fitting also results in 
  an intensity of $36.87 \pm 0.10$~mJy/beam. 
  This corresponds to a brightness temperature ($T_{\rm b}$) of $2210 \pm 6$~K. 
  We do not find a hot spot as found in the 338~GHz continuum by 
  Vlemmings et al. (\cite{vlemmings17}).

Figure~\ref{channelmap_h2o_268} shows the continuum-subtracted channel maps
of the vibrationally excited \mbox{H$_2$O}\ line
($\varv_2$ = 2, 6$_{5,2}$--7$_{4,3}$, $E_u$ = 6039.00~K) at 268.149117~GHz.
We adopted a systemic velocity (\mbox{$V_{\rm sys}$}) of
40.4~\mbox{km s$^{-1}$}\ in the local standard of rest (LSR) derived by 
Khouri et al. (\cite{khouri14}).
The relative velocity with respect to the systemic velocity
(\mbox{$V_{\rm rel}$}\ = $\mbox{$V_{\rm LSR}$} - \mbox{$V_{\rm sys}$}$) is also given in each panel. 
The continuum-subtracted channel maps reveal that 
absorption is seen over the stellar disk at \mbox{$V_{\rm rel}$}\ = $-6$ to
$0$~\mbox{km s$^{-1}$}, which is expected because of the absorption due to the cooler gas
in front of the stellar disk. 
However, the continuum-subtracted images from \mbox{$V_{\rm rel}$}\ $\approx \! -3.0$ to
$\sim$7.5~\mbox{km s$^{-1}$}\ show noticeable emission over the stellar disk, 
indicating excess emission on top of the continuum. 
In addition, prominent, spotty emission is seen along and just outside the
limb of the stellar disk at an angular distance of $\sim$40~mas
($\sim$1.9~\mbox{$R_{\star}$}), extending to $\sim$60~mas ($\sim$2.9~\mbox{$R_{\star}$}). 
The strongest emission is seen in the
east-southeast at a radius of $\sim$50~mas
(1.9~\mbox{$R_{\star}$}) at \mbox{$V_{\rm rel}$}\ = $-1.5$~\mbox{km s$^{-1}$}\ 
with an intensity of $\sim$80~mJy/beam ($T_{\rm b}$ = 4130~K). 
This is approximately
twice as strong as the continuum intensity (already subtracted in the images
shown in the figure) 
over the stellar disk, despite the high energy level of the line.

We extracted spatially resolved spectra of the \mbox{H$_2$O}\ line at four 
positions over the stellar disk and on four salient emission spots
away from the limb of the stellar disk,
as labeled in Fig.~\ref{whya_h2o_268_combined}a. 
The spectra obtained over the stellar disk
(Figs.~\ref{whya_h2o_268_combined}b--e)
show that the emission peak is redshifted by 3--5~\mbox{km s$^{-1}$}. 
However, the emission is broad, with the red wing extending to $\mbox{$V_{\rm rel}$} \approx 15$~\mbox{km s$^{-1}$}\ (position 2) 
and the blue wing extending down to $-$8~\mbox{km s$^{-1}$}\ (position 3). 
This suggests that the bulk of the material in front of the star within
$\sim$60~mas is infalling with strong turbulent motion 
and/or that the \mbox{H$_2$O}\ emission originates from layers with different
velocities (both infall and outflow) 
due to pulsation or convection with the major contribution coming from
the layers infalling at 3--5~\mbox{km s$^{-1}$}. 
  Wong et al. (\cite{wong16}) also found the presence of both infall and
  outflow in $o$~Cet observed with ALMA near minimum light. 
  We discuss the dynamics seen in our ALMA data in 
  Sect.~\ref{sect_concl}. 
The spectra extracted off the limb of the stellar disk
(Figs.~\ref{whya_h2o_268_combined}f--i)
show that the emission from positions 7 and 8 is as broad as that obtained
at positions 3 and 4. Although the emission at positions 5 and 6 is much
narrower, the spectrum at position 5 shows broad emission ranging from
  \mbox{$V_{\rm rel}$}\ = $-$8 to 8~\mbox{km s$^{-1}$}, 
and the red wing at position 6 is broader than the blue wing.

\section{Vibrationally excited \mbox{H$_2$O}\ line at 251~GHz ($\varv_2 = 2$,
  $9_{2,8}$ -- $8_{3,5}$)}
\label{sect_res_h2o_250}

We also identified -- albeit tentatively -- another vibrationally excited
\mbox{H$_2$O}\ line 
($\varv_2$ = 2, $9_{2,8}$--$8_{3,5}$, $E_u$ = 6141.05~K) at a rest frequency of
250.756830~GHz (Furtenbacher et al. \cite{furtenbacher20b}). 
The uncertainty in the rest frequency of this \mbox{H$_2$O}\ line will be discussed in greater detail in a forthcoming paper. 

Figure~\ref{channelmap_h2o_250} shows the channel maps obtained for 
this \mbox{H$_2$O}\ line
(images at $\mbox{$V_{\rm rel}$} \! > $3.0~\mbox{km s$^{-1}$}\ are not shown,
because they are affected by
the blend due to the adjacent \mbox{Si$^{17}$O}\ line at 250.744695~GHz).
While this \mbox{H$_2$O}\ line at 251~GHz and that 
at 268~GHz (Sect.~\ref{sect_res_h2o_268}) are both in the $\varv_2=2$
state with similar upper-level energies (6141 and 6039~K, respectively),
the 251~GHz \mbox{H$_2$O}\ line predominantly shows absorption over the
stellar disk,
which is in marked contrast to the prominent emission seen in the 268~GHz line across
many velocity channels. 
  The spatially resolved spectra of the 251~GHz \mbox{H$_2$O}\ line 
  show blueshifted absorption over the stellar disk
  (Fig.~\ref{whya_h2o_268_combined}, left column, blue lines), 
  with the deepest absorption blueshifted by 3--4~\mbox{km s$^{-1}$}\ at positions 1 and 2 
  (Figs.~\ref{whya_h2o_268_combined}b and c).

In the red supergiant (RSG) VY~CMa, 
Tenenbaum et al. (\cite{tenenbaum10}) reported an unidentified line 
at rest frequencies from $\sim$250.712 to 250.765~GHz, 
which is likely a blend of the lines of \mbox{$^{30}$SiO}\
($\varv$ = 2, $J$ = 6 -- 5, 250.727751~GHz),
\mbox{Si$^{17}$O}\ ($\varv=0$, $J$ = 6 -- 5, 250.744695~GHz), and
\mbox{H$_2$O}\ ($\varv_2$ = 2, $9_{2,8}$--$8_{3,5}$). 
Richards et al. (\cite{richards24}) detected the 251~GHz
\mbox{H$_2$O}\ line toward VY~CMa and 
concluded that the line is quasi-thermal.
Our observation of the (tentatively) identified 251~GHz \mbox{H$_2$O}\ line seen
in absorption over the stellar surface suggests that this is also the case for
\mbox{W~Hya}.

\section{Nonthermal origin of the 268~GHz \mbox{H$_2$O}\ line}
\label{sect_h2o_nonthermal}

Both the 251 and 268~GHz $\varv_2$ = 2 \mbox{H$_2$O}\ lines have similar upper-level
energies of $\sim$6040--6140~K. Nevertheless, the former line is only seen
in absorption.
Therefore, the excess emission in the 268 GHz line is unlikely 
caused by hot gas in front of the star, and 
the net emission over the stellar disk in the continuum-subtracted images of
the 268 GHz \mbox{H$_2$O}\ line indicates suprathermal 
($T_{\rm ex} > T_{\rm kin}$) or even maser emission ($T_{\rm ex} < 0$),
where $T_{\rm ex}$ and $T_{\rm kin}$ denote the excitation temperature and
kinetic temperature, respectively. 
Until recently there were few detections of this vibrationally excited
($\varv_2$ = 2) \mbox{H$_2$O}\ line in AGB stars and RSGs reported in the literature. 
Tenenbaum et al. (\cite{tenenbaum10}) detected a remarkably strong maser
in this line toward the RSG VY~CMa. 
Weak emission in this line was also reported in the AGB stars IK~Tau
(Velilla \mbox{Prieto} et al. \cite{velilla_prieto17}) and R~Dor
(De Beck \& \mbox{Olofsson} \cite{debeck18}), and the RSG NML~Cyg
(Singh et al. \cite{singh22}). Recently, 
Baudry et al. (\cite{baudry23}) reported the widespread presence 
of this \mbox{H$_2$O}\ line in 15 out of 17 AGB stars and RSGs observed with ALMA,
many of which show signs of maser action.

Figure~\ref{maser_model} shows the negative optical depth (maser depth)
of the 268~GHz \mbox{H$_2$O}\ line as a function of the gas temperature 
(kinetic temperature $T_{\rm kin}$) and the
ortho-\mbox{H$_2$O}\ number density at three dust temperatures (M.~Gray, priv. comm.)
based on the models of Gray et al. (\cite{gray16}) and similar to Fig.~17 of
Baudry et al. (\cite{baudry23}).
We note a change of the maser pumping regime with increasing dust
temperature. With the presence of cool dust ($T_{\rm d} \la 600$~K),
the 268~GHz \mbox{H$_2$O}\ maser is predominantly
collisionally pumped, and it requires both a rather high gas temperature 
of $\ga$1500~K and an \mbox{H$_2$O}\ density of $\ga 10^6$~cm$^{-3}$. 
With hot dust ($T_{\rm d} \ga 900$~K), maser action can occur at a lower gas
temperature ($\la\,$900 K) and \mbox{H$_2$O}\ density.
The pulsation+dust-driven models of Bladh et al. (\cite{bladh19}) show that
the density and gas temperature at 2--3~\mbox{$R_{\star}$}\ are
$10^{-14}$--$10^{-15}$~g~cm$^{-3}$ and 700--1500~K, respectively. 
These mass densities correspond to H$_2$ number densities of
$1.5 \times 10^{8}$--$1.5 \times 10^{9}$~cm$^{-3}$, which translates into
\mbox{H$_2$O}\ densities of $8 \times 10^{4}$--$8 \times 10^{5}$~cm$^{-3}$
with an ortho-\mbox{H$_2$O}\ abundance of $6 \times 10^{-4}$ 
(Khouri et al. \cite{khouri14b}).
This is consistent with the radiative pumping regime
if the dust temperature is  $\ga\,$900~K,
which is indeed the case as presented below. 
Therefore, the excess emission of the 268~GHz \mbox{H$_2$O}\ line over the
stellar disk can be due to maser amplification.
  We note that Vlemmings et al. (\cite{vlemmings21}) detected emission of
  the CO $\varv=1$ $J$ = 3 -- 2 line over the surface of W~Hya, which
  they interpreted as masers.

\section{Comparison between the 268~GHz \mbox{H$_2$O}\ $\varv_2 = 2$ emission and dust 
distribution}
\label{sect_zimpol}

\begin{figure*}
\begin{center}
\resizebox{16.8cm}{!}{\rotatebox{-90}{\includegraphics{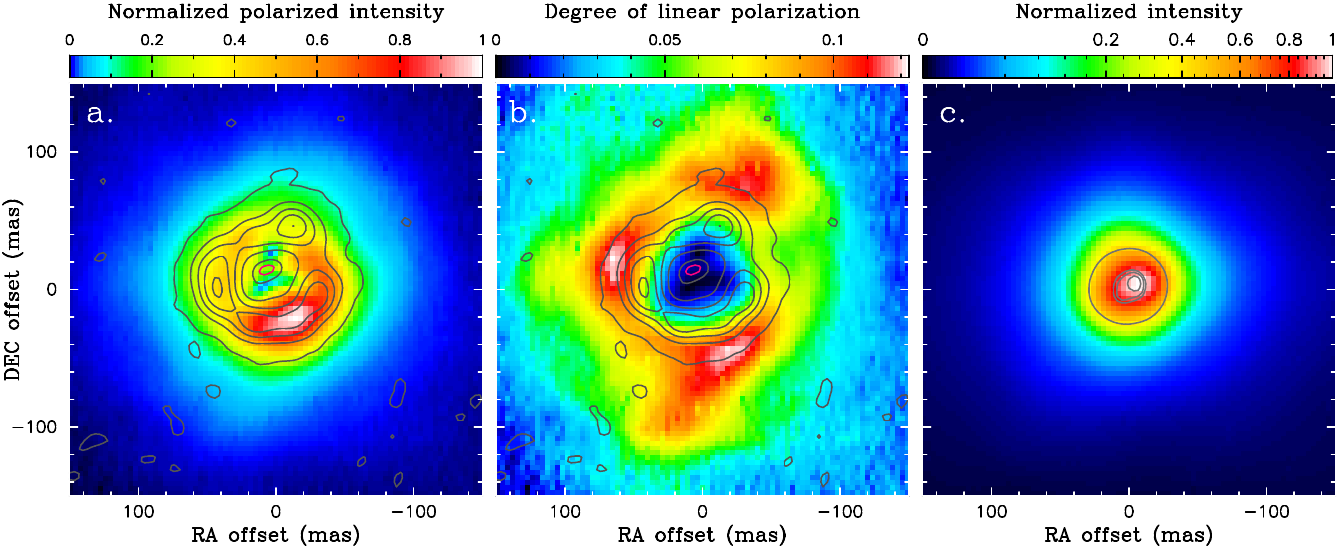}}}
\end{center}
\caption{
  Comparison between the SPHERE-ZIMPOL polarimetric imaging data and ALMA images of \mbox{W~Hya}.
  {\bf Panel (a):} Color map representing the polarized intensity map obtained at
  820~nm with a spatial resolution of 26$\times$27~mas. 
  The continuum-subtracted integrated intensity map of the 268~GHz \mbox{H$_2$O}\
  line ($\varv_2=2$, $J_{K_a,K_c}$ = $6_{5,2}$--$7_{4,3}$) obtained from
  \mbox{$V_{\rm rel}$}\ = $-1.5$ to 1.5~\mbox{km s$^{-1}$}\ is plotted with the contours.
  The dark gray contours correspond to 3, 40, 80, 120, and 160~mJy/beam~\mbox{km s$^{-1}$},
  while the magenta one corresponds to $-5$~mJy/beam~\mbox{km s$^{-1}$}. 
  {\bf Panel (b):} Color map representing the map of the degree of linear
  polarization obtained at 820~nm. The contours show the integrated
  intensity map of the 268~GHz \mbox{H$_2$O}\ line as in panel {\bf (a)}. 
  {\bf Panel (c):} Color map representing the total intensity map obtained at
  820~nm. The contours represent the 268~GHz continuum image. The outermost
  contour corresponds to 50\% of the peak intensity, which approximately
  marks the millimeter continuum stellar disk size. The inner three
  contours correspond to 97.5, 98.5, and 99.5\% of the peak intensity,
  which show the slightly offset intensity peak. 
 }
\label{zimpol_h2o_masers}
\end{figure*}

If the 268~GHz \mbox{H$_2$O}\ emission is due to maser action in the radiative
pumping regime, it traces cool, dense regions, which also provide
conditions favorable for dust to form. To examine whether there is a
correlation between the \mbox{H$_2$O}\ emission and dust formation, we compared
the 268~GHz \mbox{H$_2$O}\ emission image with the visible polarimetric imaging data
obtained with VLT/SPHERE-ZIMPOL.
We registered the SPHERE images to the ALMA image as follows. 
  Instead of 
  taking the offset intensity peak position, we fitted the SPHERE images with a uniform elliptical disk 
  (Fig.~\ref{zimpol_h2o_masers}c, color map).  
The central position of the fitted elliptical disk 
was registered to the center of the ALMA continuum image, which was also 
determined by the uniform elliptical disk fitting described in 
Sect.~\ref{sect_res_h2o_268}.

Figure~\ref{zimpol_h2o_masers}a (color map) shows the polarized intensity map
obtained at 820~nm at a spatial resolution comparable
to our ALMA images.
The polarized intensity represents the radiation scattered off dust
grains, and in an optically thin case, a polarized intensity map reflects the
spatial distribution of the dust grains. 
Figure~\ref{zimpol_h2o_masers}a  reveals clumpy dust clouds forming at a radius of $\sim$40~mas
($\sim$1.9~\mbox{$R_{\star}$}).
The map of the degree of polarization (Fig.~\ref{zimpol_h2o_masers}b, 
color map) shows clumpy dust structures extending up to $\sim$150~mas.
Three regions
with enhanced polarization of 10--12\% are seen in the north-northwest (NNW), 
east, and south. 
In both panels, the integrated intensity of the 268~GHz \mbox{H$_2$O}\ line
near the systemic velocity is shown with the contours.

Comparison between the polarized intensity map and the 268~GHz \mbox{H$_2$O}\ emission
reveals noticeable correlation. The strong \mbox{H$_2$O}\ emission 
in the south overlaps very well with the large dust cloud showing
high polarized intensity. The \mbox{H$_2$O}\ emission spots in the east and NNW also
coincide with the dust clouds. The overall shape of the \mbox{H$_2$O}\ emission with 
the NNW extension can also be recognized in the polarized
intensity map.
The previous analyses of the clumpy dust clouds forming close to the star
in \mbox{W~Hya}\ show a dust temperature of $\sim$1500~K 
(Ohnaka et al. \cite{ohnaka16}, \cite{ohnaka17}). This fulfills the high dust
temperature condition necessary for the radiatively pumped 268~GHz \mbox{H$_2$O}\
maser.
Therefore, the agreement between the \mbox{H$_2$O}\ emission and the polarization
maps is consistent with the scenario that the \mbox{H$_2$O}\ emission is of a maser
origin. 

  The total intensity map obtained with ZIMPOL is presented in
  Fig.~\ref{zimpol_h2o_masers}c. The aforementioned uniform elliptical disk
  fit results in major and minor axes of 60.6$\pm$0.3 mas and
  54.3$\pm$0.3~mas, respectively, and 
  a position angle of the major axis of 24$\pm$4\degr. 
  This size in the visible is comparable to the millimeter continuum size.
The intensity peak is offset to the northwest (NW) by
$\sim$7~mas ($\sim$0.3~\mbox{$R_{\star}$}) from the stellar disk center.
The overlaid 268~GHz continuum image (contours) also shows a slight offset 
of the intensity peak of $\sim$7~mas to the NW, although the intensity
contrast is much weaker. 
It is possible that the asymmetry in the ZIMPOL and ALMA data originates from the
same inhomogeneity in the atmosphere.
For completeness, a comparison of the \mbox{H$_2$O}\ emission with the ZIMPOL data
obtained at 748~nm is presented in Appendix~\ref{appendix_zimpol748},
which is very similar to Fig.~\ref{zimpol_h2o_masers}.

\section{Discussion and conclusion}
\label{sect_concl}

We detected unexpectedly strong nonthermal emission of the 268~GHz
\mbox{H$_2$O}\
  line covering nearly the entire surface  of the AGB star W Hya, in marked contrast to
  the thermal absorption seen in the (tentatively identified)
  251~GHz \mbox{H$_2$O}\ line. 
  The \mbox{H$_2$O}\ gas in front of the star shows infall at up to
15~\mbox{km s$^{-1}$}\ and outflow at up to $\sim$8~\mbox{km s$^{-1}$}\ 
within $\sim$2--3~\mbox{$R_{\star}$}, with the major
contribution to the emission
originating from the material infalling at 3--5~\mbox{km s$^{-1}$}. 
If the \mbox{H$_2$O}\ line emission traces dense, cool pockets as indicated
by the maser models, and the grains and gas are kinematically coupled, 
the observed spatial correlation between dust and \mbox{H$_2$O}\
indicates that the dust clouds are predominantly infalling but 
with some contribution from outflowing material.

  The dynamics within 2--3~\mbox{$R_{\star}$}\ predicted by the 1D dynamical models
  (H\"ofner et al. \cite{hoefner16}, \cite{hoefner22}) 
  is dominated by the expanding and infalling layers generated by
  stellar pulsation with velocities of up to $15$~\mbox{km s$^{-1}$}\ (infall) and 
  $\sim$10~\mbox{km s$^{-1}$}\ (outflow).
  The latest 3D models (Freytag \& H\"ofner \cite{freytag23}) show that dust
  forms and grows in large clumps, some of which are accelerated to
  $\sim$25~\mbox{km s$^{-1}$}. 
  However, they also show that there are dust grains that do not grow to
  the size sufficient for outward acceleration and that fall back to the star
  at up to $\sim$25~\mbox{km s$^{-1}$}.
  The infall and outflow velocities observed in the 268~GHz \mbox{H$_2$O}\ line 
  are in broad agreement with these theoretical models. 
  Furthermore, Liljegren et al. (\cite{liljegren17}, Fig.~5) show that
  dust grains start to form in infalling material, and they 
  are not immediately accelerated outward. This is qualitatively consistent
  with our findings (assuming the kinematical coupling of dust and gas). 
  Radiative transfer modeling of the ZIMPOL data will allow us to constrain
  the grain size, which in turn will help us estimate the radiation pressure 
  on the grains. Also, the grains that started to form in the infalling 
  material are expected to be small, which can be tested by the grain size
  derived from the modeling.

\begin{acknowledgement}
  This work is based on observations made with the Atacama Large
  Millimeter/submillimeter Array (Program ID: 2018.1.01239.S) 
  and observations collected at the European Southern Observatory under
  ESO programme 0103.D-0168(B).  
  K.O. acknowledges the support of the Agencia Nacional de 
Investigaci\'on Cient\'ifica y Desarrollo (ANID) through
the FONDECYT Regular grant 1240301. 
K.T.W. acknowledges support from the European Research Council (ERC) under
the European Union's Horizon 2020 research and innovation programme
(Grant agreement no. 883867, project EXWINGS).
We thank M.~D.~Gray for kindly providing us with the results of his maser
models. 
This research made use of the \mbox{SIMBAD} database, 
operated at the CDS, Strasbourg, France. 
\end{acknowledgement}


\begin{appendix}

\section{SPHERE-ZIMPOL observations of \mbox{W~Hya}}
\label{appendix_zimpol}

The archival SPHERE-ZIMPOL data used in our present work 
were taken on 2019 May 30 (UTC) with the filters Cnt820
(central wavelength $\lambda_c$ = 817.3~nm and full width at half
maximum (FWHM) $\Delta \lambda$ = 19.8~nm)
and Cnt748 ($\lambda_c$ = 747.4~nm and $\Delta \lambda$ = 20.6~nm). 
The star HD118877 was observed as a reference for the PSF. 
A summary of the \mbox{ZIMPOL} observations is given in
Table.~\ref{obslog_zimpol}.

We reduced the SPHERE-ZIMPOL data using the pipeline
version 0.40.0 in the same manner as described in our previous studies
(Ohnaka et al. \cite{ohnaka16}, \cite{ohnaka17}; Adam \& Ohnaka \cite{adam19}) 
and obtained the maps of total (i.e., polarized + unpolarized) intensity,
polarized intensity, and the degree of linear polarization. 
The Strehl ratios of the data of \mbox{W~Hya}\ at the observed
wavelengths were estimated from the $H$-band Strehl ratio recorded in the
auxiliary files (GEN-SPARTA data) of the SPHERE observations by applying
the Mar\'echal approximation (Adam \& Ohnaka \cite{adam19}). 
The Strehl ratios of the data of the PSF reference HD118877 were directly
measured from the observed total intensity maps. 
The derived Strehl ratios of the observations of \mbox{W~Hya}\ and HD118877 are
$\sim$0.5 and $\sim$0.3, respectively, in the visible.
The 2D Gaussian fit to the total intensity maps of the PSF reference
HD118877 resulted in PSF FWHMs of 26 $\times$ 27~mas. 

%
%


\begin{table}[h]
\caption {
  SPHERE-ZIMPOL observations of \mbox{W~Hya}\ and the PSF reference
  star HD118877.
}
\begin{center}
  \tabcolsep = 3pt
\begin{tabular}{l c l l l l l}\hline
Object & $t_{\rm obs}$ & Filter     & Seeing    & AM & Strehl & Strehl \\
   & (UTC)       & (cam1) & (\arcsec) &   & ($H$) & (Cnt820) \\
   &             & (cam2) &           &   &       & (Cnt748) \\
\hline
\multicolumn{7}{c}{2019 May 30 (UTC)}\\
\hline
\mbox{W~Hya} & 01:05:13 & Cnt820   & 0.72 & 1.0 & 0.85 & 0.51 \\
             &          & Cnt748   &      &     &      & 0.48 \\
HD118877 & 01:47:41 & Cnt820 & 0.74 & 1.0 & 0.68 & 0.32 \\
         &          & Cnt748 &      &     &      & 0.26 \\
\hline
\label{obslog_zimpol}
\vspace*{-7mm}

\end{tabular}
\end{center}
\tablefoot{
AM: Airmass. 
The Strehl ratios in the visible were computed from the $H$-band Strehl ratios 
for \mbox{W~Hya}, while they were measured from the observed ZIMPOL images for 
HD118877. 
}
\end{table}

\section{Millimeter continuum image of \mbox{W~Hya}}

\begin{figure}[h]
\begin{center}
\resizebox{\hsize}{!}{\rotatebox{0}{\includegraphics{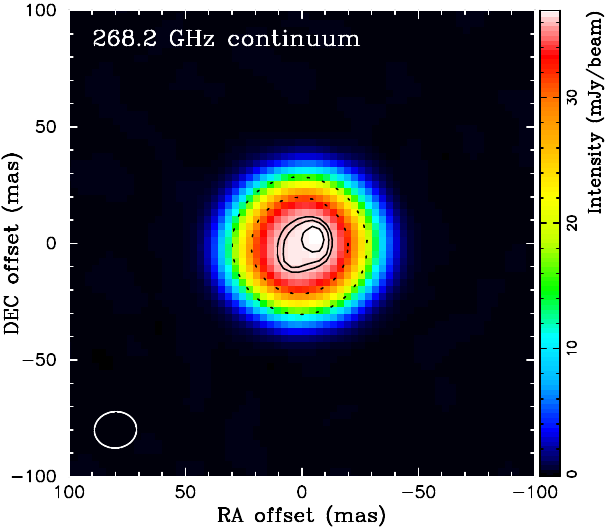}}}
\end{center}
\caption{
  Continuum image of \mbox{W~Hya}\ obtained in the spectral window centered
  at 268.2~GHz with a restoring beam of $18 \times 16$~mas. 
  The contours correspond to 97.5, 98.5, and 99.5\% of the maximum
  intensity and show the slight offset of the intensity peak.
  The outer black dotted line represents the elliptical fit to the image
    (59.1 $\times$ 57.7~mas),
  while the inner black dotted circle represents the stellar angular diameter
  of 40.7~mas obtained in the near-infrared.
}
\label{cont_image}
\end{figure}

Figure~\ref{cont_image} shows the image of \mbox{W~Hya}\ obtained in the 268~GHz
continuum with a restoring beam of 18 $\times$ 16~mas.
The angular diameter of $\sim$60~mas obtained by fitting with a uniform
elliptical disk is $\sim$1.4 times larger than the stellar angular
diameter of 40.7~mas. 
The intensity peak is slightly offset by $\sim$7~mas ($\sim$0.34~\mbox{$R_{\star}$})
to the NW from the disk center.
However, the asymmetry in the intensity is 1\% at most, as plotted with the
contours in Fig.~\ref{cont_image}.

\vspace*{3cm}

\section{Spatially resolved spectra of the $\varv_2=2$ \mbox{H$_2$O}\ lines
}
\label{appendex_h2o_combined}

\begin{figure*}[h]
\begin{center}
\resizebox{16.3cm}{!}{\rotatebox{0}{\includegraphics{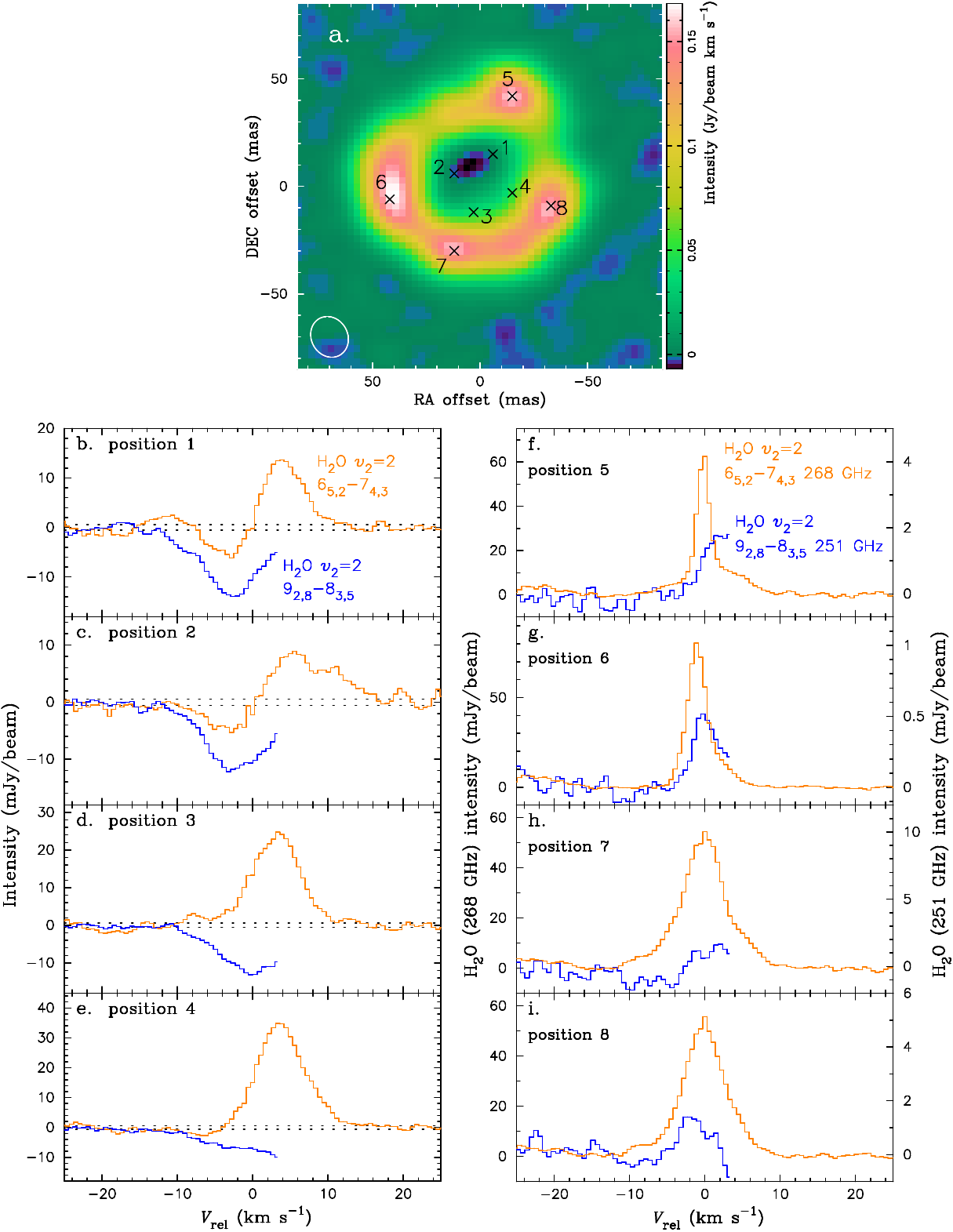}}}
\end{center}
\caption{
  Spatially resolved spectra of two vibrationally excited \mbox{H$_2$O}\ lines 
  detected toward \mbox{W~Hya}. 
  {\bf Panel (a):} The continuum-subtracted intensity map of the 268~GHz \mbox{H$_2$O}\ line
  ($\varv_2=2$, 6$_{5,2}$--7$_{4,3}$) integrated from \mbox{$V_{\rm rel}$}\ = $-1.5$ to
  1.5~\mbox{km s$^{-1}$}\ to show the most prominent emission components. 
  The crosses and numbers correspond to the positions where the spatially
  resolved spectra shown in the lower two columns were obtained 
    (the spectra were extracted at the marked pixels).
  {\bf Panels (b)}--{\bf (i):}
    Spatially resolved spectra obtained at positions 1--4
    over the stellar disk and positions 5--8 off the
    limb of the stellar disk. 
    The orange and blue lines show the continuum-subtracted spectra of the
    \mbox{H$_2$O}\ lines $\varv_2=2$, 6$_{5,2}$--7$_{4,3}$ (268~GHz) and
    $\varv_2=2$, 9$_{2,8}$--8$_{3,5}$ (251~GHz), respectively. 
    In panels {\bf (f)}--{\bf (i)},
    the left and right ordinates correspond to the
    intensity of the lines at 268 and 251~GHz, respectively.
  The spectra of the $\varv_2=2$, 9$_{2,8}$--8$_{3,5}$ line are shown
  only up to \mbox{$V_{\rm rel}$}\ = 3.2~\mbox{km s$^{-1}$}, because the data 
  at more redshifted
  velocities are affected by the blend due to the adjacent \mbox{Si$^{17}$O}\
  line at 250.744695~GHz. 
    The weak emission seen at $V_{\rm rel} \! < \! -20$~\mbox{km s$^{-1}$}\ in the
    268~GHz line in panels {\bf (f)}--{\bf (i)} is due to the
    SO$_2$ ($\varv_2=1$, $11_{3,9}$--$11_{2,10}$) line at 268.169791~GHz. 
    The dotted lines in panels {\bf (b)}--{\bf (e)} mark the RMS noise level.
}
\label{whya_h2o_268_combined}
\end{figure*}

Figure~\ref{whya_h2o_268_combined} shows the spatially resolved spectra
of the \mbox{H$_2$O}\ lines $\varv_2=2$, 6$_{5,2}$--7$_{4,3}$ and
$\varv_2=2$, 9$_{2,8}$--8$_{3,5}$ extracted from the continuum-subtracted
data.

\FloatBarrier

\section{268~GHz \mbox{H$_2$O}\ maser conditions}
\label{appendix_masers}

\begin{wrapfigure}{l}{183mm}
\vspace*{-100mm}
\begin{center}
\resizebox{8.5cm}{!}{\rotatebox{0}{\includegraphics{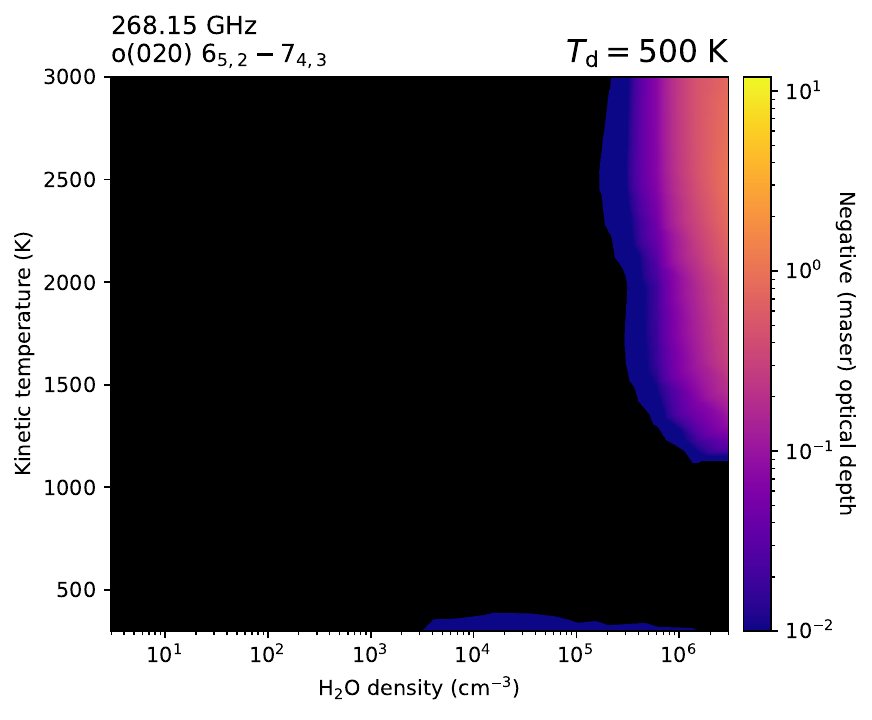}}}
\resizebox{8.5cm}{!}{\rotatebox{0}{\includegraphics{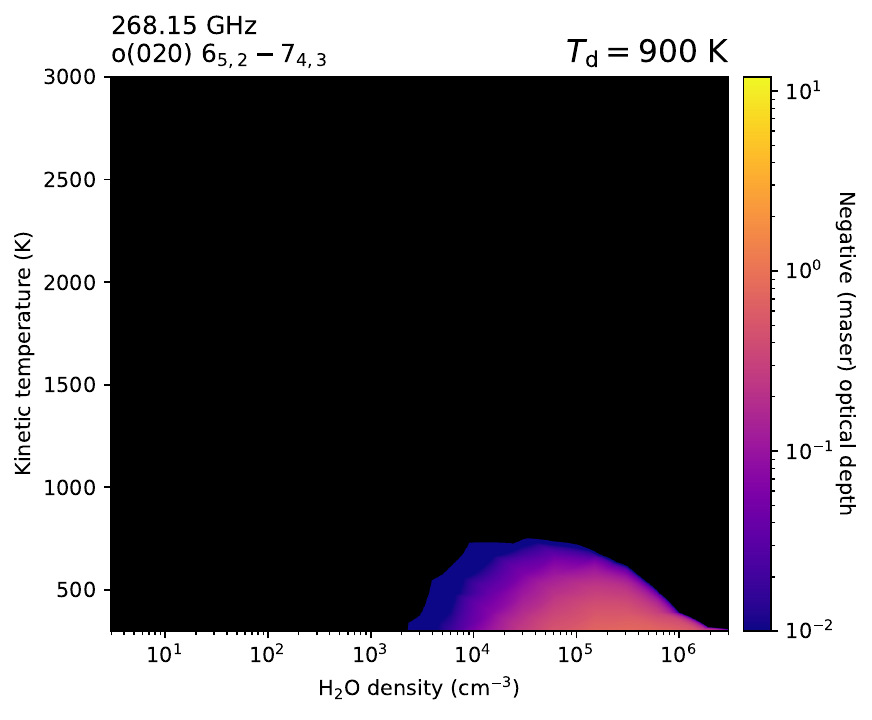}}}
\resizebox{8.5cm}{!}{\rotatebox{0}{\includegraphics{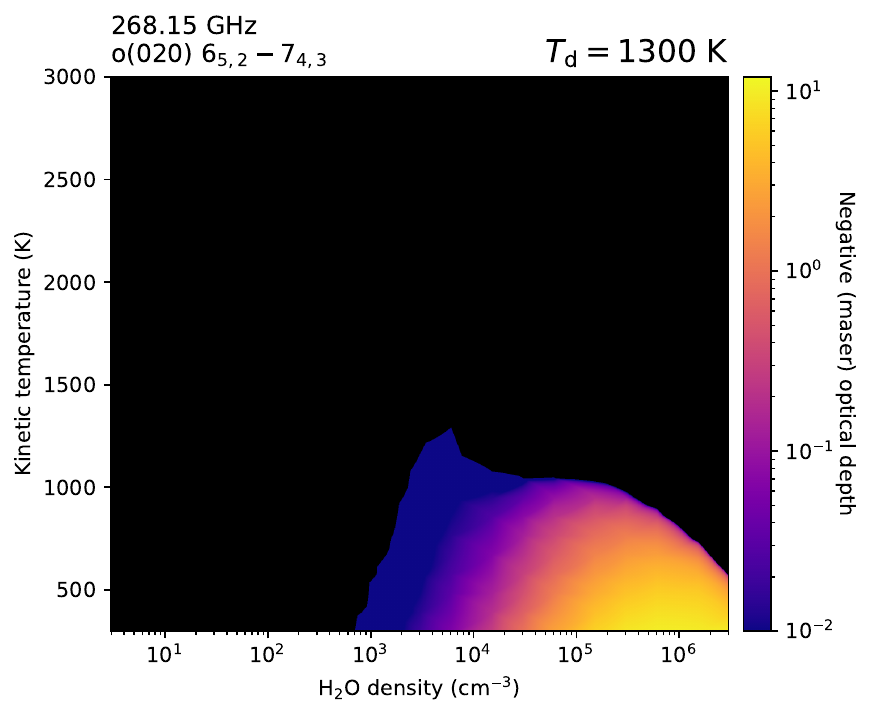}}}
\end{center}
\caption{
Negative optical depth (maser depth) of the 268~GHz \mbox{H$_2$O}\ line 
($\varv_2$ = 2, 6$_{5,2}$--7$_{4,3}$) as a function of
the gas temperature (kinetic temperature) and the ortho-\mbox{H$_2$O}\ number density
at dust temperatures ($T_{\rm d}$) of 500, 900, and 1300~K. 
}
\label{maser_model}
\end{wrapfigure}

Figure~\ref{maser_model} shows the negative optical depth of the
268~GHz \mbox{H$_2$O}\ line ($\varv_2$ = 2, 6$_{5,2}$--7$_{4,3}$) as a function of
the gas temperature (\mbox{kinetic} temperature) and the ortho-\mbox{H$_2$O}\
number
density at three different dust temperatures based on the model of Gray et al.
(\cite{gray16}). 
\vspace*{10cm}

\clearpage

\section{Comparison of the ALMA image of the 268~GHz \mbox{H$_2$O}\ $\varv_2=2$ line
with the ZIMPOL data obtained at 748~nm}
\label{appendix_zimpol748}

\begin{wrapfigure}{l}{183mm}
\begin{center}
\resizebox{17cm}{!}{\rotatebox{-90}{\includegraphics{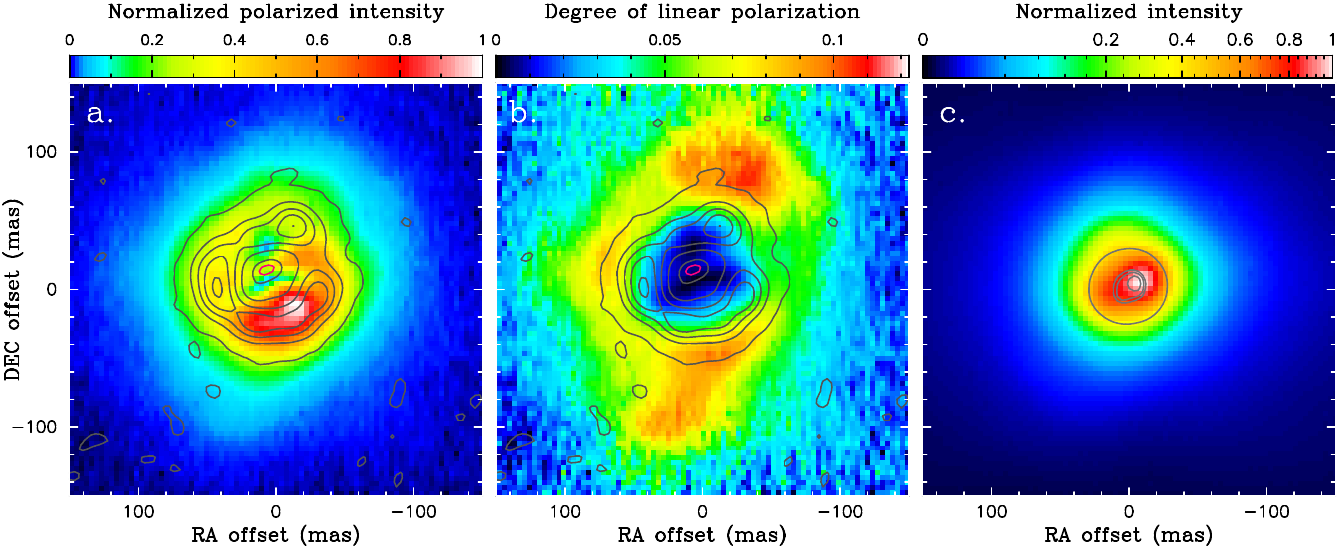}}}
\end{center}
\caption{
  Comparison between the SPHERE-ZIMPOL polarimetric imaging data obtained
  at 748~nm and ALMA images of the 268~GHz \mbox{H$_2$O}\ line 
  ($\varv_2$ = 2, $J_{K_a,K_c}$ = $6_{5,2}$--$7_{4,3}$)
  toward \mbox{W~Hya}, shown in the same manner as Fig.~\ref{zimpol_h2o_masers}.
 }
\label{zimpol748_h2o_masers}
\end{wrapfigure}

Figure~\ref{zimpol748_h2o_masers} shows a comparison of the ALMA image of
the 268~GHz \mbox{H$_2$O}\ line ($\varv_2$ = 2, $J_{K_a,K_c}$ = $6_{5,2}$--$7_{4,3}$)
with the ZIMPOL data obtained at 748~nm. The results are very similar to
those shown in Fig.~\ref{zimpol_h2o_masers}. The degree of polarization
at 748~nm is slightly lower than that measured at 820~nm.
\vspace*{5mm}

\end{appendix}

\end{document}